\documentclass[aps,twocolumn,floatfix]{revtex4}

\usepackage{graphicx}
\usepackage{dcolumn}
\usepackage{bm}

\newcommand{\pdiffl}[2]{\frac{\partial #1}{\partial #2}}

\begin{document}


\title{Non-Iterative Characteristics Analysis \\ for High-Pressure Ramp Loading}

\date{January 20, 2015, revisions to October 1, 2018 -- LLNL-JRNL-704258}

\author{Damian C. Swift}
\email{dswift@llnl.gov}
\author{Dayne E. Fratanduono}
\author{Richard G. Kraus}
\author{Evan A. Dowling\footnote{%
Current affiliation: University of Maryland --College Park 
}}
\affiliation{%
   Physics Division, Lawrence Livermore National Laboratory,
   7000 East Avenue, Livermore, California 94551, USA
}

\begin{abstract}
In the canonical ramp compression experiment, a smoothly-increasing load is
applied to the surface of the sample, and the particle velocity history is measured
at two or more different distances into the sample, at interfaces
where the surface of the sample can be probed. The velocity
histories are used to deduce a stress-density relation, usually using 
iterative Lagrangian analysis to account for the
perturbing effect of the impedance mismatch at the interface. In that technique,
a stress-density relation is assumed in order to correct for the perturbation, 
and is adjusted until it becomes consistent with the deduced stress-density relation.
This process is subject to the usual difficulties of nonlinear
optimization, such as the existence of local minima (sensitivity to the initial
guess), possible failure to converge, and relatively large computational effort.
We show that, by considering the interaction of successive characteristics
reaching the interfaces, the stress-density relation can be deduced directly by
recursion rather than iteration.  This calculation is orders of magnitude faster
than iterative analysis, and does not require an initial guess. Direct
recursion may be less suitable for very noisy data, but it was robust when
applied to trial data. The stress-density relation deduced was identical to the
result from iterative Lagrangian analysis.
\end{abstract}


\maketitle

\section{Introduction}
Ramp loading is of increasing importance for the study of matter at high
pressures, where it allows the compressibility to be measured continuously
along a trajectory close to an isentrope (i.e. with less heating than induced
by a shock wave) in a single dynamic loading experiment
\cite{Hall2001}.
Samples can be subjected to high pressures at much lower temperatures than in
shock wave experiments, making it possible to explore solid phases over 
a wider pressure range and producing states much closer to planetary isentropes.
From the difference between shock and ramp loading states, the
thermal contributions to the equation of state (EOS) can be deduced,
represented for example by the Gr\"uneisen parameter.

Ramp loading has been achieved using a variety of techniques, the first
identifiable method being reloading following
the expansion of product gases from chemical explosives \cite{Barnes1974}.
Projectile impacts, the canonical method for inducing shocks,
have been modified for ramp loading by using an impactor
constructed so that the impedance increased with distance from the
impact surface \cite{Barker1984}.
For low-impedance samples, isentropic compression has been induced
approximately by sandwiching the sample between high-impedance anvils
in which a shock is introduced; the sample rings up toward the
incident shock pressure by a succession of weaker shocks \cite{Nellis1996}.
Ramps as a practical method to probe compression states in between
shocks and isotherms were investigated more actively with the development
of pulsed magnetic field loading \cite{Hall2001}.
Since then, loading techniques have grown to include
the reloading following the expansion of matter
shocked above the critical isentrope by a laser pulse \cite{Edwards2004},
ablation by a temporally-shaped laser pulse \cite{Swift_lice_2005},
reloading following the
expansion of matter heated by a laser-driven hohlraum \cite{Smith2007},
and direct ablation by hohlraum radiation heated with temporally-shaped
laser pulses \cite{Bradley2009}.

Although often close to an isentrope, ramp loading is not strctly
isentropic because of irreversible processes occurring in uniaxial loading,
such as plastic flow.
Given additional information, such as knowledge about the constitutive
behavior of the sample material, the isentrope can be deduced from
ramp measurememnts \cite{Kraus2016}.
For convenience, unless this point is relevant, we refer below to the 
thermodynamic trajectory associated with the ramp as an isentrope.

As we discuss in more detail below, ramp-loading data has generally been 
analyzed using
an iterative technique in which the isentrope is assumed, and corrections
are made repeatedly using the ramp data until the
modified isentrope converges \cite{Rothman2006}.
With such iterative refinement techniques, there is always a potential
concern that the solution found may depend on the isentrope initially assumed,
or on the parameters of the iterative scheme.
Here we present a non-iterative analysis method that we find gives the same
result more stably with less computational effort.

\section{Isentrope measurement from ramp loading}
\begin{figure}
\begin{center}\includegraphics[scale=0.72]{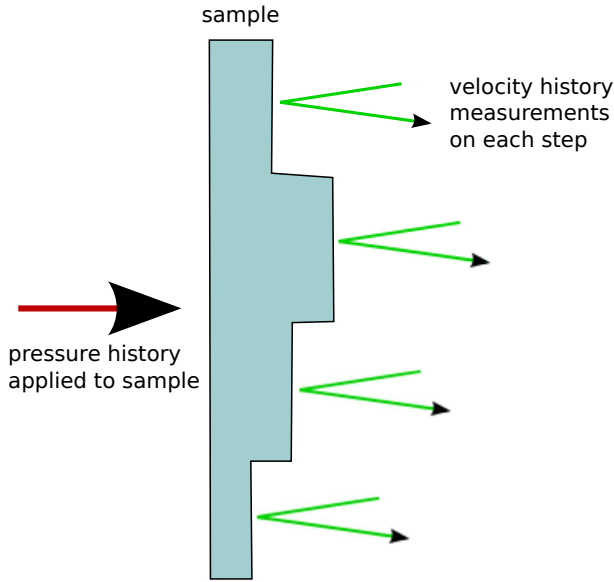}\end{center}
\caption{Schematic of ramp wave EOS experiment.}
\label{fig:rampschem}
\end{figure}

Ramp-loading experiments can be configured to measure the stress-density
response of the sample material by observing the ramp after it has
propagated through different thicknesses of material.
For matter of a given composition in a single thermodynamic phase,
the sound speed increases with compression, so the higher-pressure tail
of a ramp catches up with the lower-pressure head, leading to a steepening
of the ramp which can be measured and analyzed.
The most mature diagnostic method is to measure the material velocity
history $u(t)$ at each of a set of steps of different thickness $x$
(Fig.~\ref{fig:rampschem}).
We wish to deduce the propagation speed $c$ of the characteristic as a function
of the particle speed $u_p$.
If the presence of the steps did not perturb the ramp, $c(u_p)$ could be
deduced directly from the velocity history at each step.
However, the step is an impedance mismatch, and introduces perturbing 
characteristics that interact with subsequent characteristics 
emanating from the ramp drive, so comparing the time at which each step
reaches a given velocity gives only an apparent characteristic velocity,
\begin{equation}
c'(u_p)=\Delta x/\Delta t(u_p;x),
\end{equation}
which must be corrected to deduce $c(u_p)$.
$c$ is the longitudinal sound speed with respect to uncompressed material,
which, ignoring any contribution from elastic shear stress, 
is related to the sound speed calculated from
the EOS, $c_e=c\rho_0/\rho$ where $\rho$ is the mass density
and $\rho_0$ its initial value.
Given $c(u_p)$, one can therefore calculate the normal stress $p$ 
(the pressure, in the absence of shear stresses) as a function of $\rho$ using
Riemann integrals,
\begin{equation}
p=\int \rho c_e\,du_p; \quad \rho=\int \frac{\rho}{c_e}du_p
\end{equation}
where the EOS sound speed is
\begin{equation}
c_e\equiv\sqrt{B_s/\rho}
\end{equation}
and $B_s$ is the isentropic bulk modulus, 
$\left.\rho\,\partial p/\partial\rho\right|_s$.

In the iterative analysis used previously \cite{Rothman2006},
an estimate of the isentrope $\tilde p_0(\rho)$ is used
to correct the apparent characteristic speed $c'(u_p)$ and thus deduce
a modified estimate of the isentrope $\tilde p_1(\rho)$,
repeating until $\tilde p_i(\rho)$ converges.
This is in effect a multivariate non-linear fitting process,
which can be computationally intensive if $\tilde p_i(\rho)$ is slow to
converge, can potentially fail by locating a local minimum or can potentially
yield solutions which depend on the initial guess $\tilde p_0(\rho)$,
and which is potentially prone to numerical instabilities when the
solutions $\tilde p_i(\rho)$ oscillate around the actual isentrope
with increasing amplitude.

As we show next, it is not necessary to assume the solution and
iteratively improve it, as all the information is present for a deterministic
solution.

\section{Ramp characteristics at an impedance mismatch}
\begin{figure}
\begin{center}\includegraphics[scale=0.72]{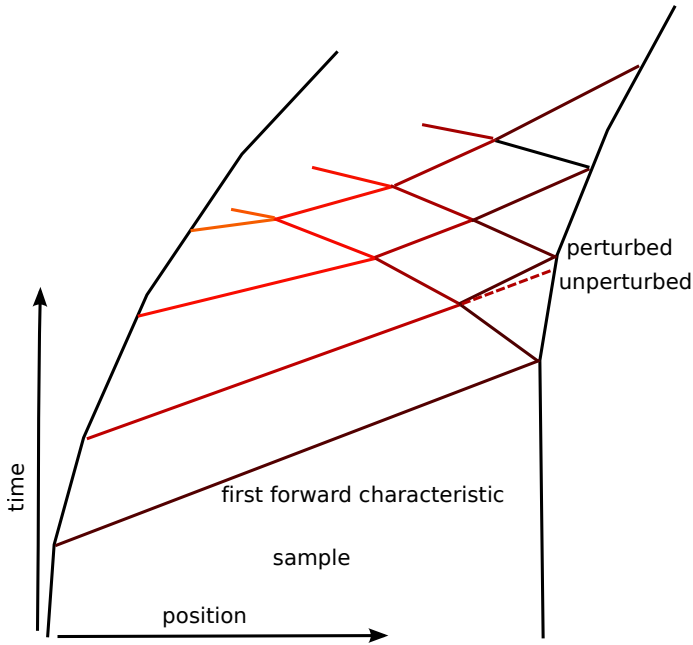}\end{center}
\caption{Interaction between successive reflected characteristics
   in the laboratory frame.}
\label{fig:characs}
\end{figure}

\begin{figure}
\begin{center}\includegraphics[scale=0.72]{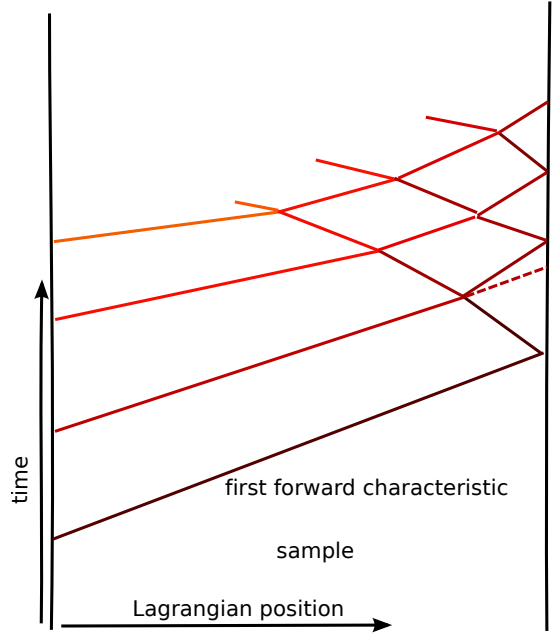}\end{center}
\caption{Interaction between successive reflected characteristics
   in the Lagrangian frame.}
\label{fig:characslag}
\end{figure}

Consider a ramp load applied to one side of a sample of some thickness.
At each pressure, the load propagates through the sample as a
characteristic of some speed $c$.
As the characteristic approaches the opposite side of the sample,
it is perturbed by interactions with the reflection of preceding
characteristics from the free surface, releasing the pressure gradually to
zero at the surface (Fig.~\ref{fig:characs}).
When calculating intersections between the characteristics, it is easier to
work in the Lagrangian frame, where distances are expressed in terms of the
original position of each element of matter within the sample
(Fig.~\ref{fig:characslag}). The front and back surfaces are then constant
in time, and wave speeds are scaled by the compression $\rho/\rho_0$.

At the interface,
any given characteristic has been affected by all previous (lower-pressure)
characteristics, but not by any subsequent (higher pressure) ones.
The effect can be corrected for recursively.
Consider the appearance of the characteristic $\chi_1$ resulting in the first
significant change in interface velocity at some step 
of thickness $x$ (Fig.~\ref{fig:characs}).
This characteristic is unaffected by previous characteristics, so its speed $c_1$ can be determined
trivially by the difference in arrival time at steps of different height.

The next characteristic $\chi_2$ appears at some later time $t_2$.
However, its propagation to the interface has been perturbed by the reflection of $\chi_1$, which travels backward at $-c_1$.
As the speed of this reflected wave is known, the location at which it must have interacted with the unperturbed $\chi_2$
can be calculated:
\begin{eqnarray}
t_2^\prime & = & \frac{x_1-x_2}{2 c_1}+\frac{t_1+t_2}2 \\
x_2^\prime & = & x_2 - c_1 (t_2^\prime - t_2)
\end{eqnarray}
This correction can be made for each step, and the corresponding correction
can be made for all subsequent characteristics.

Once a characteristic has been corrected for all preceding characteristics, its Lagrangian wave speed
(here taken to mean with respect to uncompressed material) can be calculated from the corrected position-time data
for two or more steps: $c_i = \Delta x_i^\prime/\Delta t_i^\prime$.
The step height only enters as a difference, so absolute step heights are
irrelevant so long as the relative heights are known.

In previous studies of the nature of ramp wave analysis, there was some
discussion of the inherent nature of the problem \cite{Hinch2010,Ockendon2010}.
According to our analysis, the determination of an isentrope
from surface velocities is inherently deterministic.
It is an inverse problem as normally stated, but is well-posed and invertible.
A similar approach was postulated in connection with the analysis of ramp wave
data using the hodograph transform \cite{Ockendon2010}, although 
as a brief aside, and does not appear to have been developed further.

Although our recursive but deterministic algorithm seems potentially 
advantageous for speed and to remove the need to guess an approximate
isentrope, there is the potential for poor practical performance
such as sensitivity to noise.
We investigated the performance of the algorithm on simulated experimental data,
for which the isentrope could be calculated independently,
as described in Section~\ref{sec:test} below.

\section{Calculation of compression and normal stress}
For a free surface, and ignoring irreversible effects such as plastic flow,
the particle speed $u_p$ corresponding to a given interface speed $u$ is $u/2$.
The mass density then changes with particle speed as
\begin{equation}
\pdiffl\rho{u_p} = \frac\rho{c_e}.
\end{equation}
The mass density can be integrated implicitly through the characteristics, which is generally more accurate than
a simple finite difference update:
\begin{equation}
\rho_{i+1} = \rho_i \frac{1+\Delta u_p/2 \bar c_e}{1-\Delta u_p/2 \bar c_e}
\end{equation}
where $\Delta u_p$ is the difference in inferred particle velocity between characteristics $i$ and $i+1$,
and $\bar c_e$ is the average longitudinal wave speed.

The corresponding rate of change of pressure through the characteristics is
\begin{equation}
\pdiffl p{u_p} = \rho c_e
\end{equation}
which can be integrated numerically as
\begin{equation}
p_{i+1} = p_i + \Delta u_p \bar\rho \bar c_e.
\end{equation}

\begin{figure*}
\begin{center}\includegraphics[scale=0.45]{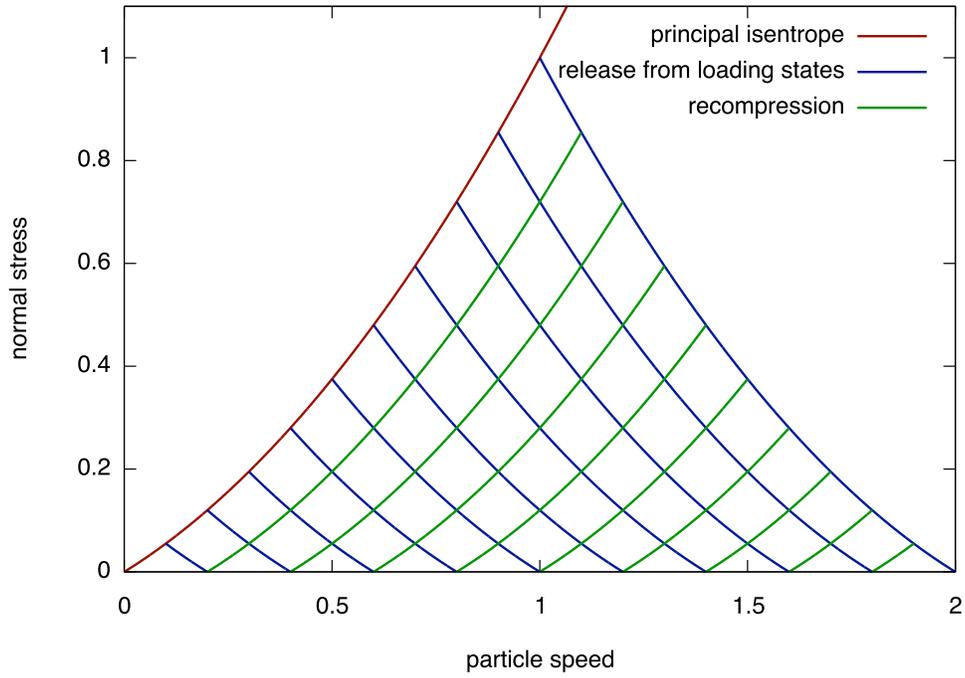}\end{center}
\caption{Pressure-particle speed states occurring with free-surface velocity
   sampled at uniformly-spaced intervals.}
\label{fig:puequal}
\end{figure*}

\begin{figure*}
\begin{center}\includegraphics[scale=0.45]{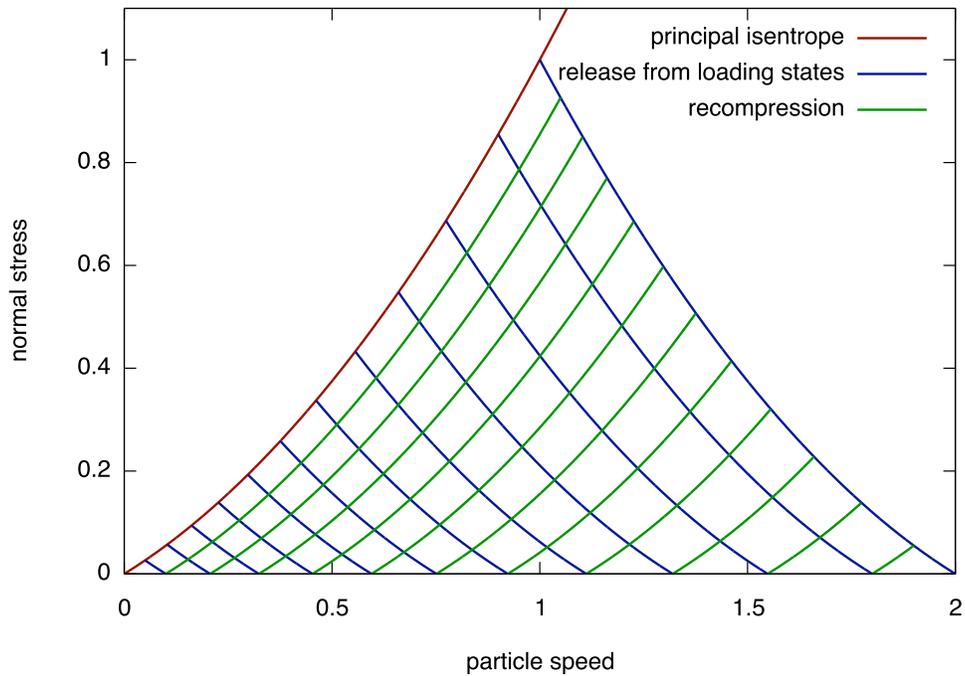}\end{center}
\caption{Pressure-particle speed states occurring with free-surface velocity
   sampled at geometrically-varying intervals.}
\label{fig:pugeom}
\end{figure*}

This finite difference analysis assumes that the sound speed is constant
between the points at which the $i$th forward characteristic
intersects the $j$th backward characteristic,
i.e.
\begin{equation}
c_{i,i-1}=c_{i+1,i}.
\end{equation}
Considering the sequence of states induced by the
interacting characteristics in pressure-particle speed space,
the speeds are equal if characteristics are inferred from
surface velocities interpolated at uniform intervals
(Fig.~\ref{fig:puequal}): intersections between $N$ forward
and $N$ backward characteristics occur at $N$ pressures,
corresponding to the pressures deduced along the isentrope.
In contrast, if velocities are analyzed at varying intervals
-- which occurs naturally in experimental data taken at uniform
intervals of time -- then the intersections between characteristics
occur at intermediate pressures, so interpolation is required in order to
perform the recursive analysis (Fig.~\ref{fig:pugeom}).

\section{Analysis of simulated data}\label{sec:test}
\begin{figure*}
\begin{center}\includegraphics[scale=0.45]{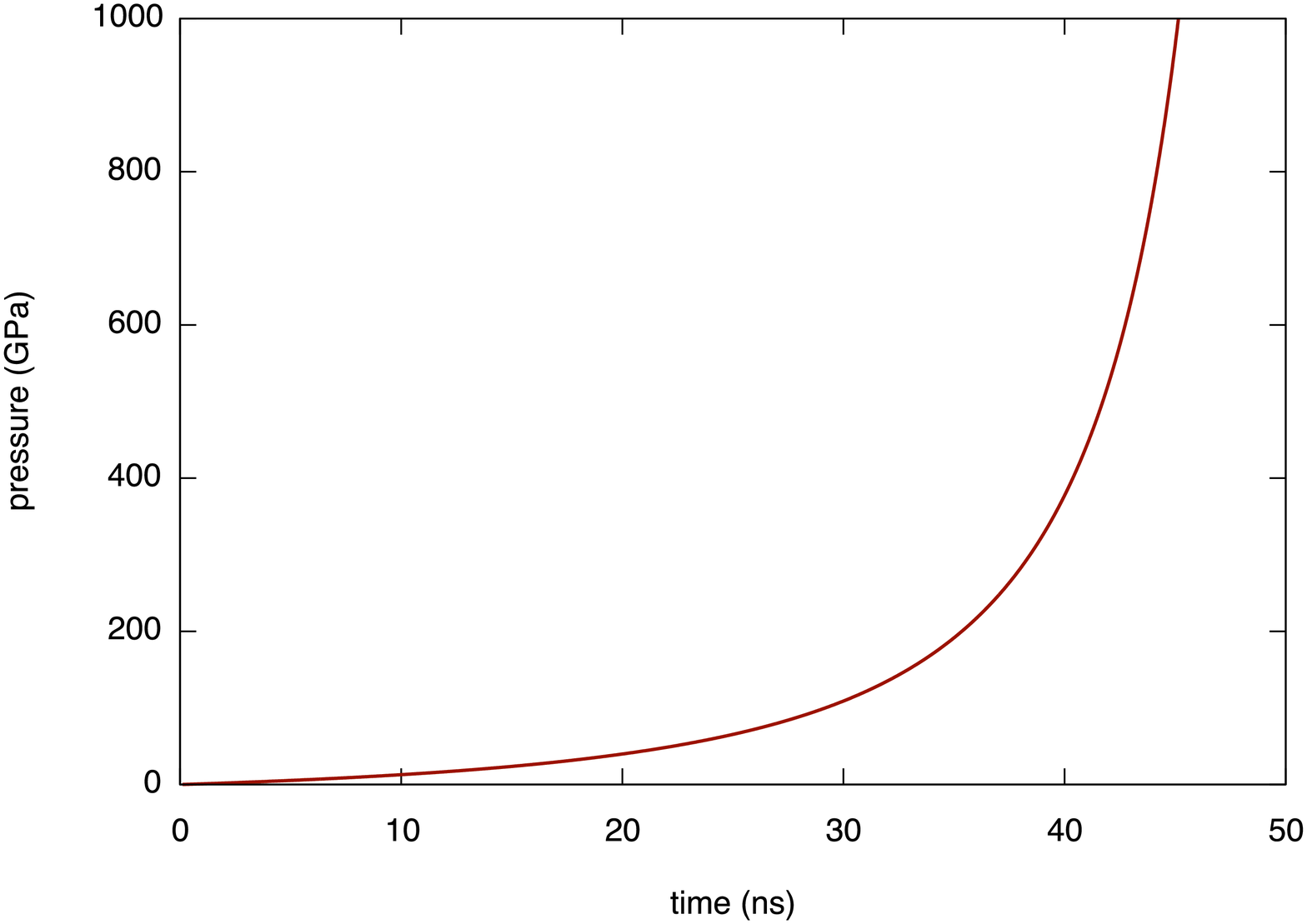}\end{center}
\caption{Loading history applied to the Cu to generate simulated data.}
\label{fig:load}
\end{figure*}

\begin{figure*}
\begin{center}\includegraphics[scale=0.45]{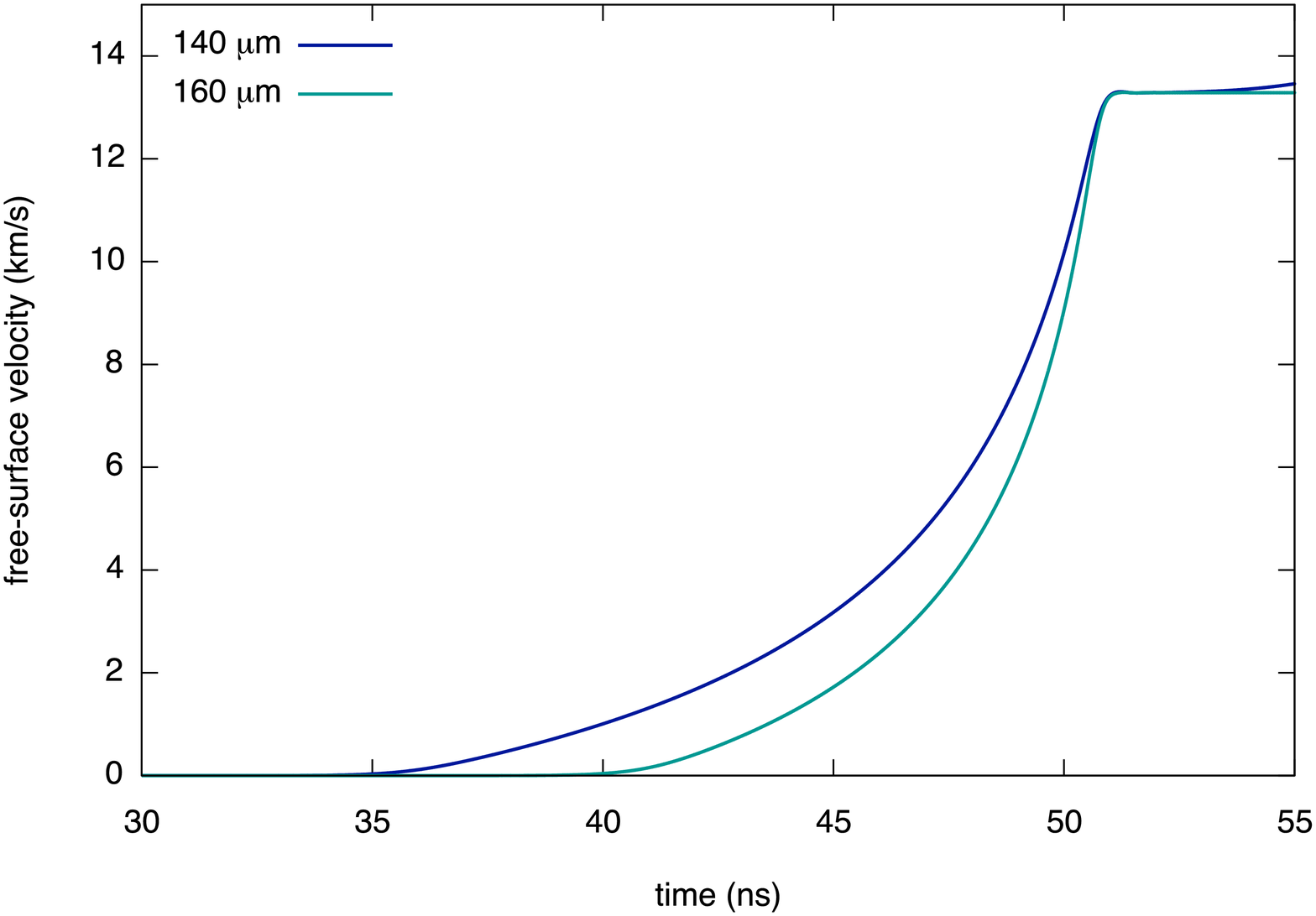}\end{center}
\caption{Simulated free-surface velocity histories for steps of different
   thickness, showing steepening of ramp wave.}
\label{fig:velhist}
\end{figure*}

As a test case, we consider the analysis of simulated data for Cu
ramp-loaded to 1\,TPa.
Cu was treated as behaving according to an analytic EOS of the 
Gr\"uneisen form \cite{Steinberg1996}.
The peak pressure represents a compression of 2.25,  making the loading history 
prone to forming a shock if not carefully designed.
The loading history was chosen to be the ideal shape \cite{Swift_idealramp_2008}
for an isentrope from this EOS as calculated by numerical integration
\cite{Swift_genscalar_2008} -- the ideal shape being one where all the 
characteristics cross to form a shock at the same point in the material.
This choice does not affect the ramp analysis, but it does eliminate
trial and error in choosing step heights suitable for the analysis.
The ideal shape was scaled so that the shock formation distance was
200\,$\mu$m (Fig.~\ref{fig:load}),
with suitable steps being 140 and 160\,$\mu$m.
Simulated experimental data were generated from continuum mechanics simulations 
of the velocity history at the surface of each step (Fig.~\ref{fig:velhist}).
The simulations were performed using a Lagrangian hydrocode
with a second-order predictor-corrector numerical scheme
using artificial viscosity to stabilize the flow against unphysical 
oscillations \cite{Benson1992,LAGJ}.
Simulations were performed with spatial resolutions of 1 to 0.1\,$\mu$m.
Gaussian noise was added to the data in time or velocity.

\begin{figure*}
\begin{center}\includegraphics[scale=1.0]{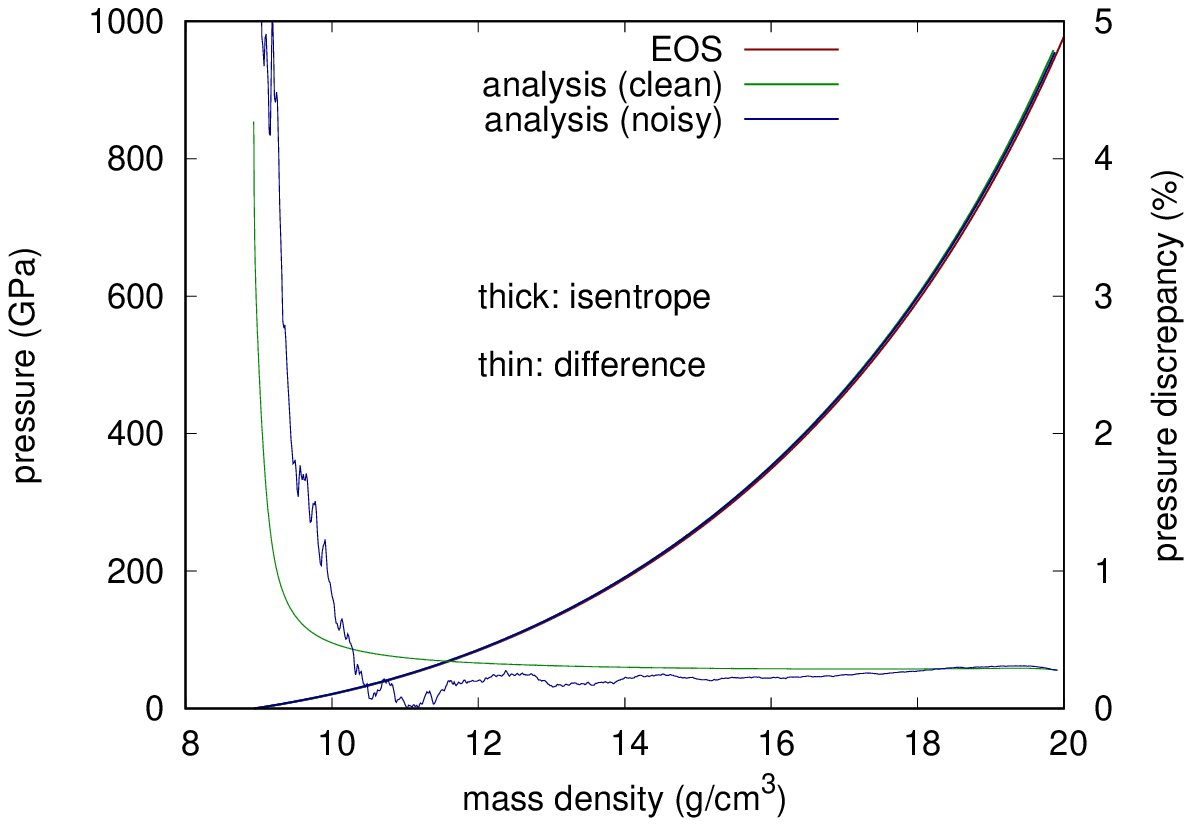}\end{center}
\caption{Isentrope obtained by direct integration and by recursive 
   characteristics analysis.
   (The analysis of data with noise added lies between the isentrope
   integrated directly from the EOS and the analysis of clean data.)}
\label{fig:analysis}
\end{figure*}

The recursive analysis method was implemented as a Java program,
and applied to the simulated data, with and without noise.
Without noise, the algorithm reproduced the $p(\rho)$ isentrope as calculated 
directly by integration of the EOS to within 0.5\%\ of the pressure
at any given mass density.
The discrepancy reduced when the simulated data were generated with
finer spatial resolution or sampled with finer temporal resolution, 
implying that the inaccuracy was dominated by
the precision of the hydrocode simulations rather than the analysis method.
With noise, the analysis algorithm experienced numerical difficulties if
any tabulated $u(t)$ was non-monotonic, which is a trivial filtering
constraint. Once filtered, the expected isentrope was recovered to an accuracy
commensurate with the noise level.
(Fig.~\ref{fig:analysis}.)

We did not conduct a rigorous comparison of the computational speed
of the recursive algorithm with the iterative algorithm, partly because
the latter exhibited difficulties converging to the degree
and with the robustness of the former -- we found cases where
the iterative algorithm failed to converge to a solution, whereas
the recursive algorithm continued to reproduce the correct solution.
For a typical experimental dataset, with the time sampled at a few hundred
values of velocity, the iterative algorithm typically took a few minutes
to converge to a solution.
The recursive algorithm took around 10\,ms, i.e. around $3\times 10^4$
faster.
(The execution time was dominated by the $\sim$1\,s taken to read the
ASCII dataset.)
For a high resolution dataset, with several thousand points from each step,
the increase in speed was approximately $2\times 10^5$.
Efficiency of this order are useful to make it practicable to embed the
ramp analysis within another process, such as iteration over parameters in
an additional model such as time-dependence, or uncertainty analysis by Monte-Carlo
perturbation of parameters or experimental data.
High-throughput analysis also makes it possible to reduce ramp data on times
commensurate with the rate of data acquisition on high repetition-rate 
(multi-Hertz, or faster) experimental platforms.

\section{Conclusions}
We have developed a non-iterative algorithm -- deterministic, and based
on recursion over characteristics -- for analyzing ramp-loading data,
which gives the same results as the iterative algorithm generally used.
No numerical problems were found in processing data with noise,
though the data had to be filtered to be monotonic.

Comparing with simulated data representative of real experiments,
the recursive algorithm was several orders of magnitude faster than
the iterative algorithm, and performed more stably, giving a solution
in cases where the iterative algorithm failed to converge.
Unlike an iterative solution, the recursive algorithm does not require
an estimate of the solution to be made as a starting condition, which
is undesirable as it may bias the solution.

The deduction of a stress-density relation for a material from surface
velocity measurements is inherently deterministic.
It is an inverse problem as usually stated, but appears well-posed
and invertible.

\section{Acknowledgments}
Jon Eggert provided an implementation of the iterative algorithm.
Steve Rothman (AWE) drew our attention to the papers by Ockendon and Hinch.
Tom Arsenlis, Jim McNaney, and Dennis McNabb provided funding and encouragement
via the National Nuclear Security Agency's Science Campaign.
This work was performed under the auspices of the
U.S. Department of Energy by Lawrence Livermore National Laboratory
under Contract DE-AC52-07NA27344.

\end{document}